\documentclass[aps,prl,superscriptaddress,twocolumn,floatfix]{revtex4-2}
\usepackage{units}
\usepackage{amsmath}
\usepackage{amssymb}
\usepackage{graphicx}
\usepackage{bm}
\usepackage{multirow,color,relsize,ulem,microtype}
    
\newcommand{\cc}[1]{{\color{red}#1}}
\newcommand{\nc}[1]{{\color{blue}#1}}
\newcommand{\be}{\begin{equation}}
\newcommand{\ee}{\end{equation}}
\newcommand{\re}[1]{\text{Re}[#1]}
\newcommand{\im}[1]{\text{Im}[#1]}

\begin{document}

\title{Cyclic ladder operators and hidden Weyl-Heisenberg structure in a Floquet system}

\author{Tianao Wu}
\affiliation{Department of Physics and Astronomy, College of Staten Island, CUNY, Staten Island, NY, USA}
\affiliation{The Graduate Center, CUNY, New York, NY, USA}

\author{Li Ge}
 \email{li.ge@csi.cuny.edu}
\affiliation{Department of Physics and Astronomy, College of Staten Island, CUNY, Staten Island, NY, USA}
\affiliation{The Graduate Center, CUNY, New York, NY, USA}

\begin{abstract}
Ladder operators, found in the quantum harmonic oscillator and other quantized systems, provide an elegant approach to solving or understanding otherwise intricate physics problems. In this Letter, we discuss cyclic ladder operators in both Hermitian and non-Hermitian systems with a finite Hilbert space, with the highest (lowest) level directly descending (ascending) to the lowest (highest) level via a single raising (lowering) operation. We show that an equally spaced energy ladder emerges when these systems have an underlying Weyl-Heisenberg commutation relation, with the cyclic ladder operators and the temporal evolution operator behaving as the generators of the Weyl-Heisenberg group. We further illustrate such a system using a one-dimensional Floquet lattice, where the cyclic ladder operators become diagonal and the temporal evolution simplifies to a permutation matrix after a Floquet period. Our findings reveal a hidden relation between non-trivial dynamics and algebraic principles in Floquet systems, which may exist for other quantum numbers as well besides the energy levels.
\end{abstract}

\maketitle
In the early days of quantum mechanics, the non-commutative multiplication rule of position states 
suggested by Heisenberg led 
Born and 
Jordan to usher in the matrix mechanics \cite{Born}, and it was 
Weyl who first realized that 
\be
[\hat{x},\hat{p}] =i\hbar \label{eq:Born}
\ee 
implies an underlying mathematical structure with unitary elements $U(x,p)=e^{i(p\hat{x}-x\hat{p})}$ \cite{Weyl_book}, which is now known as the Weyl-Heisenberg (WH) group. 

This group was a key step in establishing the equivalence of the matrix mechanics and the wave mechanics, via an extension of the Stone-von Neumann theorem \cite{Stone,vonNeumann}. It is also the governing group of the displacement operator of coherent states in quantum optics \cite{glauber_coherent_1963}. When the continuous observables (e.g., $x$ and $p$) are replaced by discrete quantities, the WH group can be characterized by its three generators $\nu$, $X$ and $Z$, satisfying 
\be
\nu XZ = ZX,\quad \nu = e^{2\pi i/N}\,(N\in\mathbb{Z}). \label{eq:WH}
\ee 
$X$ and $Z$ correspond to two elements $U(x_1,p_1),U(x_2,p_2)$ in the continuous case. 
Such discrete WH groups were recently utilized in the study of informationally complete quantum measurements \cite{dariano_informationally_2004,medendorp_experimental_2011,tabia_experimental_2012}, where $X$ can assume the simple form of a permutation on different components of the wave function, and $Z$ is given by a diagonal operator that shifts their phase progressively by $\nu^n\,(n=0,1,\ldots,N-1)$. 

The commutation relation (\ref{eq:Born}), when applied to the quantum harmonic oscillator (QHO), naturally leads to the ladder operators: 
\be 
a |\psi_{n+1}\rangle = |\psi_n\rangle, \quad a^\dagger |\psi_{n}\rangle = |\psi_{n+1}\rangle, \label{eq:1}
\ee 
where $|\psi_n\rangle$'s are the eigenstates of the Hamiltonian $H$ with discrete energy levels $E_n=n\Delta\geq 0\,(n=0,1,\ldots)$. This equally spaced energy ladder, however, is not warranted by the commutation relation (\ref{eq:Born}) or (\ref{eq:WH}). For example, adding anharmonic terms to $H$ distorts the ladder. Instead, this equally spaced energy ladder is a direct consequence of a different relation, i.e.,
\be 
[H,a] = -a\Delta. \label{eq:2}
\ee 
Note that this relation holds for an infinite Hilbert space with the energy unbounded from above.  

In this Letter, we show that equally spaced energy ladders in a \textit{finite-dimensional} Hilbert space \textit{all} emerge from the WH commutation relation (\ref{eq:WH}) in the absence of degeneracy, even in a non-Hermitian system \cite{feng2017non} with a complex energy ladder. 

We exemplify this finding using Floquet (i.e., time-modulated) engineering, a powerful tool to construct effective quantum dynamics and spectral structures beyond those accessible in static systems \cite{rudner_band_2020}. By applying a periodic drive in both classical \cite{rechtsman_photonic_2013,mukherjee_observation_2020,chitsazi2017experimental,hockendorf2019non} and quantum systems \cite{Oka,Kitagawa,Lindner,potirniche_floquet_2017}, Floquet protocols enable the realization of a wide range of effective models and unconventional spectral features, including engineered quasienergy dispersions and dynamical band structures with no equilibrium counterparts. 

Beyond their phenomenological richness, the spectral structures emerging in Floquet systems often signal the presence of hidden algebraic principles governing the underlying dynamics. In our case, we identify an equally spaced energy ladder when the Floquet bandgap is closed. Motivated by this observation, we uncover the algebraic origin of this structure and demonstrate that its ladder operators are generators of the WH group.\\

\noindent \textit{Equally spaced energy ladder}--We start by noting the similarity of the permutation operator $X$ in Eq.~(\ref{eq:WH}) and the ladder operators in the energy eigenbasis in a finite Hilbert space. With $N$ non-degenerate energy levels, we choose $X$ such that 
\be
X|\psi_{n+1}\rangle = |\psi_n\rangle, \quad X^\dagger|\psi_n\rangle = |\psi_{n+1}\rangle
\label{eq:cLadder0}
\ee 
with $|\psi_{n+N}\rangle = |\psi_n\rangle$. In other words, we find that $X,X^\dagger$ behave as standard ladder operators except for
\be
X|\psi_0\rangle = |\psi_{N-1}\rangle,\quad X^\dagger |\psi_{N-1}\rangle = |\psi_0\rangle. \label{eq:cLadder}
\ee 
We will refer to these operators as cyclic ladder operators and denote them again by $a,a^\dagger$. 

Next, we wish to establish a commutation relation similar to Eq.~(\ref{eq:2}) that warrants an equally spaced energy ladder $E_n=n\Delta\geq 0$. Because $H$ is diagonal in the energy eigenbasis, it is easy to show that while Eq.~(\ref{eq:2}) no longer holds, it can be remedied by defining two additional operators $b,b^\dagger$ such that
\be 
[H,a] = -(a + b)\Delta,\quad [a^\dagger,H] =  -(a^\dagger + b^\dagger)\Delta. \label{eq:cLadder2}
\ee 
$b$ and $b^\dagger$ satisfy 
\be 
b|\psi_n\rangle = b^\dagger|\psi_{n+1}\rangle = 0 \,\,(n=1,\ldots,N-1),\label{eq:b1}
\ee 
\be 
b|\psi_0\rangle = -N |\psi_{N-1}\rangle, \;\; b^\dagger|\psi_{N-1}\rangle = -N |\psi_0\rangle\label{eq:b2},
\ee 
and we will refer to them as the  ``escalator'' operators. 
When applied to $|\psi_n\rangle\,(n>0)$, the first relation in Eq.~(\ref{eq:cLadder2}) leads to 
\be
H a|\psi_n\rangle = [aH-a\Delta]|\psi_n\rangle = (E_n-\Delta)a|\psi_n\rangle, \nonumber 
\ee
i.e., $a|\psi_n\rangle$ is a lower energy level at energy $E_n-\Delta$. Thus we have an energy ladder spaced by $\Delta$, similar to the QHO. 

To show that Eq.~(\ref{eq:cLadder2}) indeed satisfies the cyclic property (\ref{eq:cLadder}), next we denote $a|\psi_0\rangle = |\phi\rangle$ and find
\be
H |\phi\rangle = [aH-(a+b)\Delta]|\psi_0\rangle = (E_0-\Delta)|\phi\rangle + N\Delta |\psi_{N-1}\rangle\nonumber
\ee
using Eqs.~(\ref{eq:cLadder2})--(\ref{eq:b2}). Multiplying $\langle\psi_n|$ to the result above from the left, we find
\be
E_n \langle\psi_n|\phi\rangle = (E_0 - \Delta) \langle\psi_n|\phi\rangle + N\Delta\, \delta_{n,N-1}
\ee
or simply 
\be
(n+1) \langle\psi_n|\phi\rangle = N \,\delta_{n,N-1},
\ee
i.e., $a|\psi_0\rangle = |\phi\rangle = |\psi_{N-1}\rangle$. The other relation $a^\dagger|\psi_{N-1}\rangle = |\psi_0\rangle$ in Eq.~(\ref{eq:cLadder}) can be similarly proven using the second relation in Eq.~(\ref{eq:cLadder2}). 

Having shown that Eq.~(\ref{eq:cLadder2}) warrants an equally spaced real energy ladder, below we explore if the ladder can exist in a non-Hermitian system \cite{feng2017non} as well with a complex $\Delta$. One may argue that the two relations in Eq.~(\ref{eq:cLadder2}) hold simultaneously because they are Hermitian conjugates, which requires $H$ to be Hermitian and $\Delta$ real. By lifting them, it may then seem unlikely to have an equally spaced energy ladder in a non-Hermitian system. 

We show, however, the two relations in Eq.~(\ref{eq:cLadder2}) coexist even when $H$ is non-Hermitian and $\Delta$ is complex. This behavior is possible due to the commutative property of the cyclic ladder operators, i.e.,
\be 
a^\dagger a = a a^\dagger = 1,\label{eq:com}
\ee  
and in addition, that $a^\dagger b$ is Hermitian. By multiplying the first relation in Eq.~(\ref{eq:cLadder2}) by $a^\dagger$ from left \textit{and} right, we then find
\be
[a^\dagger,H] = -(a^\dagger + a^\dagger b a^\dagger)\Delta = -(a^\dagger + b^\dagger a a^\dagger)\Delta,  \label{eq:cL}
\ee
which gives the second relation in Eq.~(\ref{eq:cLadder2}). \\

\noindent \textit{WH commutation relation}--Now let us establish the connection between the WH commutation relation (\ref{eq:WH}) and the equally spaced energy ladder. Again, $X$ is one cyclic ladder operator, and clearly $H$ does not play the role of $Z$ [see Eq.~(\ref{eq:cLadder2})]. 

We note, however, the WH commutation relation emerges when we use the temporal evolution operator $U(t) = e^{-iHt}$ ($\hbar\equiv 1$) as our $Z$: In the energy eigenbasis, $U(t)$ is diagonal with elements $u_n = e^{-in\Delta t}$, and using the commutation relation (\ref{eq:cLadder2}), it can be shown that 
\be
\nu \,a U(t) = U(t) a   \label{eq:WH2}
\ee
hold at $t=2\pi/(N\Delta)\equiv T$ (see End Matter), where $U(t)$ and its integer powers form a cyclic group themselves, i.e., $U(T)^N=1$.  

This observation points to the direction where such an energy ladder can be realized beyond a time-independent Hamiltonian, i.e., in a Floquet system. $U(T)^N=U(NT)=1$ indicates that an arbitrary state $|\psi(0)\rangle=\sum_n c_n |\psi_n\rangle$ returns to itself after a period of $NT$. This behavior is guaranteed when the Hamiltonian is time-independent (such as in a $J_x$ lattice \cite{christandl_perfect_2004,wolterink_supersymmetric_2023} and Wannier-Stark lattice \cite{longhi_bloch_2009,xu_experimental_2016}):
\be
|\psi(NT)\rangle = \sum_n e^{-in\Delta NT} c_n |\psi_n\rangle = \sum_n c_n |\psi_n\rangle = |\psi(0)\rangle.\nonumber
\ee 
And it can also take place in a Floquet system, where $c_n$ is now time-dependent but periodic in the Floquet period $T'$, i.e., $c_n(t) = c_n(t+T')$. We then find
\be
|\psi(NT)\rangle = \sum_n e^{-in\Delta NT} c_n(NT) |\psi_n\rangle = \sum_n c_n(NT) |\psi_n\rangle\nonumber
\ee 
equals $|\psi(0)\rangle$ if $c_n(NT)=c_n(0)$, i.e., $NT$ is an integer time of $T'$ as well, which can be satisfied simply by $T=T'$. This is the case in the example we will explore later.

The WH commutation relation (\ref{eq:WH2}) with a real $\Delta$ also implies a particularly simple dynamical behavior after just $T$ (instead of $NT$). The key observation is the symmetry between $a$ and $U(t)$ in Eq.~(\ref{eq:WH2}) [and $X,Z$ in Eq.~(\ref{eq:WH})]: Eq.~(\ref{eq:WH2}) still holds when we exchange $a$ and $U(t)$, with $u_1^{-1}$ also replaced by $u_1$. In the energy eigenbasis, our cyclic ladder operator $a$ is a permutation matrix while the temporal evolution $U(t)$ is diagonal. In the diagonal basis of $a$, we then expect that $U(t)$ becomes a permutation matrix at some time $t$. 

To confirm this hypothesis, we first note that the commutative relation (\ref{eq:com}), which is absent for standard ladder operators, indicates that our cyclic ladder operators $a,a^\dagger$ can be diagonalized simultaneously. To see that $a$ and $a^\dagger$ share the same eigenstates in the energy eigenbasis, we note that they are both circulant matrices \cite{davis_circulant_1979} and that all circulant matrices of the same size share the same eigenbasis, independent of their matrix elements:
\begin{gather} 
|\phi_n) = \frac{1}{\sqrt{N}}[1,\nu^n,\nu^{2n},\ldots,\nu^{(N-1)n}].\label{eq:planewave}
\end{gather}
Here $\nu$ is again $e^{2\pi i/N}$ and $n=0,1,\ldots,N-1$. We have used the ``round'' bra-ket notation because $a$ is non-Hermitian \cite{feng2017non} as a permutation matrix, and hence both a right eigenbasis \{$\phi_n$\} and a left eigenbasis \{$\tilde{\phi}_n$\} are needed to diagonalize it:
\be 
a |\phi_n) = a_n |\phi_n),\;\; (\tilde{\phi}_n| a = a_n (\tilde{\phi}_n|,\;\;a_n = \nu^{-n}.
\ee  
Here the row vector $(\tilde{\phi}_n|$ is the transpose of the column vector $|\tilde{\phi}_n)$. 
By choosing $|\tilde{\phi}_0)=|\phi_0)$ and $|\tilde{\phi}_n)=|\phi_{N-n})\,(n=1,2,\ldots,N-2)$, the following biorthogonal relation holds for the unconjugated inner product between the \{$\phi_n$\} and \{$\tilde{\phi}_n$\}:
\be 
(\tilde{\phi}_{n'}|\phi_n) = (\phi_n|\tilde{\phi}_{n'}) = \delta_{nn'}.
\ee 
In this biorthogonal basis, the cyclic ladder operators are indeed diagonal, and we denote them by  
\be 
d = \text{Diag}(1,\nu^{-1},\ldots,\nu^{1-N}),\quad d^\dagger = d^*\label{eq:a_altBasis}
\ee 
to distinguish them from their permutation forms (i.e., $a,a^\dagger$) in the energy eigenbasis. 

The Hamiltonian now takes a Toeplitz form (i.e., diagonal-constant), with 
\be 
H_{nn'} = (\tilde{\phi}_n|H|\phi_{n'}) = \frac{\Delta}{\nu^{n-n'}-1}\;\;(n\neq n')\label{eq:H_altBasis}
\ee 
and $H_{nn}={\sum_n E_n}/{N}=(N-1)\Delta/2$. 
We note that $H$ is Hermitian only when $\Delta$ is real, but its energy levels always form a ladder with the energy spacing $\Delta$.


Such an $H$, when viewed as a lattice model, is fairly complicated with long-range couplings, and one would not have anticipated that it has an equally spaced energy spectrum. The general form of the temporal evolution operator $U(t)$ is also a full matrix, i.e.,
\be 
U_{nn'}(t) =\frac{1-e^{-iN\Delta t}}{N(1-\tilde{\nu})},\;\,\tilde{\nu}\equiv e^{i(\Delta t + 2\pi \frac{n'-n}{N})}. \label{eq:U}
\ee  
However, the symmetry between $a$ and $U(t)$ implied by the WH commutation relation (\ref{eq:WH2}) noted previously manifests itself when $t=T=2\pi/(N\Delta)$, at which the numerator in $U_{nn'}(t)$ vanishes. Meanwhile, the denominator in $U_{nn'}(t)$ is finite except for $U_{n,n-1}(t)\,(n>0)$ and $U_{0,N-1}$, with which we find
\be
U(T) = a^\dagger \label{eq:Uperm}
\ee
and $U(T)^N=1$. $a^\dagger$ is again the permutation form of the cyclic ladder operator in the energy eigenbasis. We exemplify this dynamical behavior and the equally spaced energy ladder next in a Floquet system, where the cyclic ladder operators become diagonal in position space. \\

\begin{figure}[b]
\centering
\includegraphics[width=\linewidth]{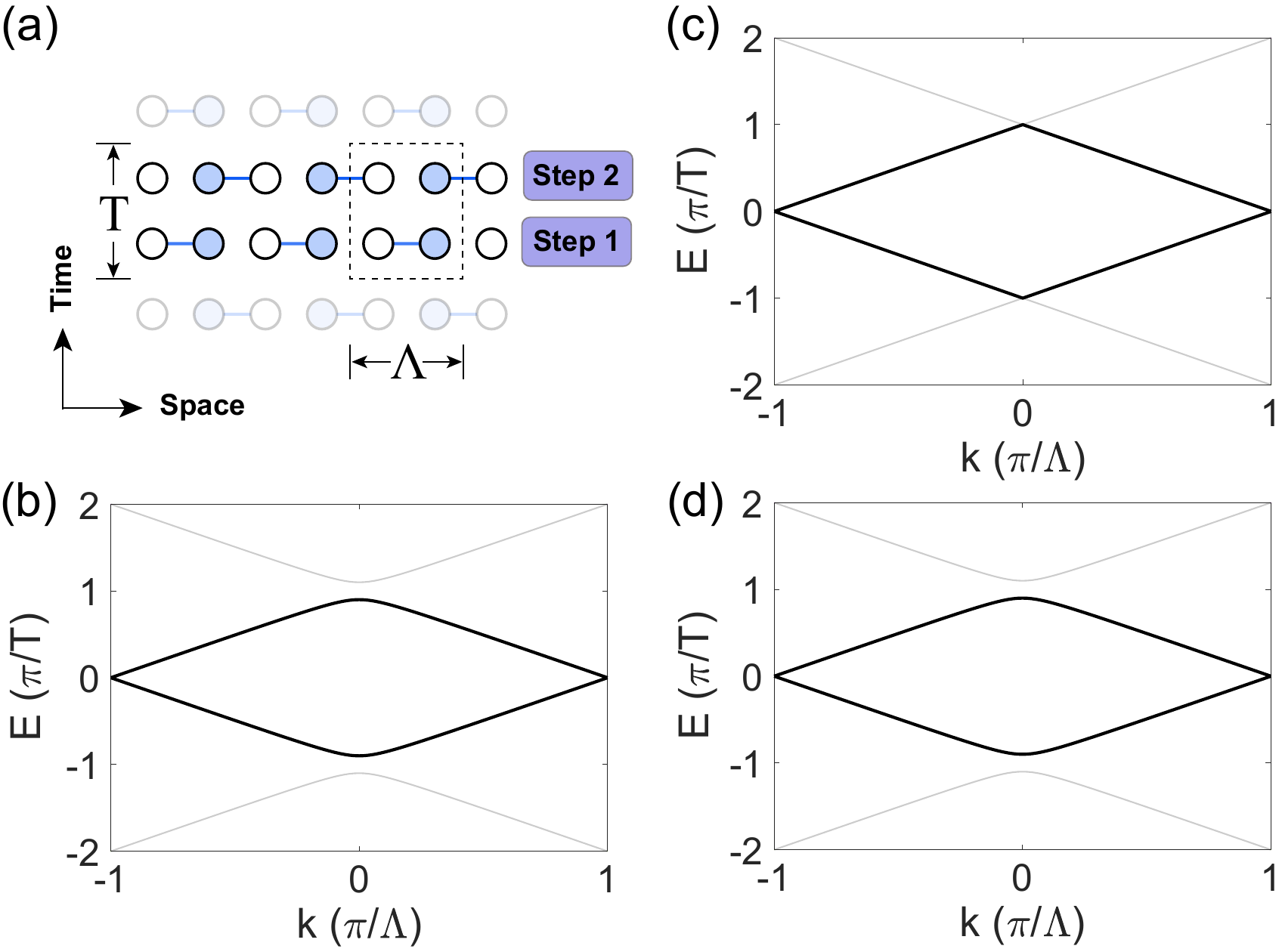}
\caption{(a) Schematics of a one-dimensional Floquet system with temporal period $T$ and spatial period $\Lambda$. Couplings are marked by bars, and one unit cell is boxed by the dashed line. Sublattices A and B are marked by filled and open dots. (b-d) Band structure of the Floquet system with $g=1$ and $T/\pi=0.9,1,1.1$, respectively. The first Floquet BZ is darkened. 
}
\label{fig:1}
\end{figure}

\noindent \textit{Floquet system}--Consider the tight-binding model in Fig.\ref{fig:1}(a), which is a one-dimensional version of the system studied in Ref. \cite{rudner_anomalous_2013}. In it, nearest-neighbor couplings are applied alternately in two steps, each with time $T/2$. We denote their respective temporal evolutions by $U_j = e^{-ih_jT/2}\,(j=1,2)$, where the static Hamiltonians $h_{1,2}$ with the periodic boundary condition are given by
\be 
h_1 = g\sigma_x,\quad h_2 = g( \sigma_x \cos k\Lambda - \sigma_y \sin k\Lambda )\equiv g\tilde{\sigma}_x.
\ee 
$g$ is the coupling, $k$ is the lattice momentum, $\Lambda$ is the spatial period, and we have taken $\hbar=1$. The temporal evolution in a Floquet period $T$ is then given by $U(T) = U_2U_1 \equiv e^{-iH_\text{eff}T}$, which has the form
\be 
U(T) = \cos^2 \theta - i(\sigma_x+\tilde{\sigma}_x)\cos \theta \sin\theta - \sigma_x\tilde{\sigma}_x\sin^2\theta
\ee 
with $\theta=gT/2$. The band structure of this two-band Floquet Hamiltonian $H_\text{eff}$ is shown in Figs.~\ref{fig:1}(b--d) for different values of $\theta$. It captures a single Floquet Brillouin zone (BZ), and the same band structure repeats itself vertically with a $2\pi/T$ spacing. 

The band gaps below and above the two bands in the first BZ close at $E = \pm\pi/T$ with $\cos\theta=0$. With this vanishing cosine, the temporal evolution operator becomes
\be
U(T) = -\sigma_x\tilde{\sigma}_x = -
\begin{pmatrix}
e^{-ik\Lambda} & \\
& e^{ik\Lambda} 
\end{pmatrix}, 
\ee 
from which we find a diagonal 
\be
H_\text{eff} =  
\begin{bmatrix}
(|k|\Lambda - \pi)/T & \\
& (\pi - |k|\Lambda)/T 
\end{bmatrix} 
\ee
in the first BZ. Consequently, a Dirac cone is formed at $E = \pm\pi/T$, with linear dispersion (i.e., $d \omega/d k=\pm \Lambda/T$) throughout the BZ [Fig.~\ref{fig:1}(c)].

\begin{figure}[t]
\centering
\includegraphics[width=\linewidth]{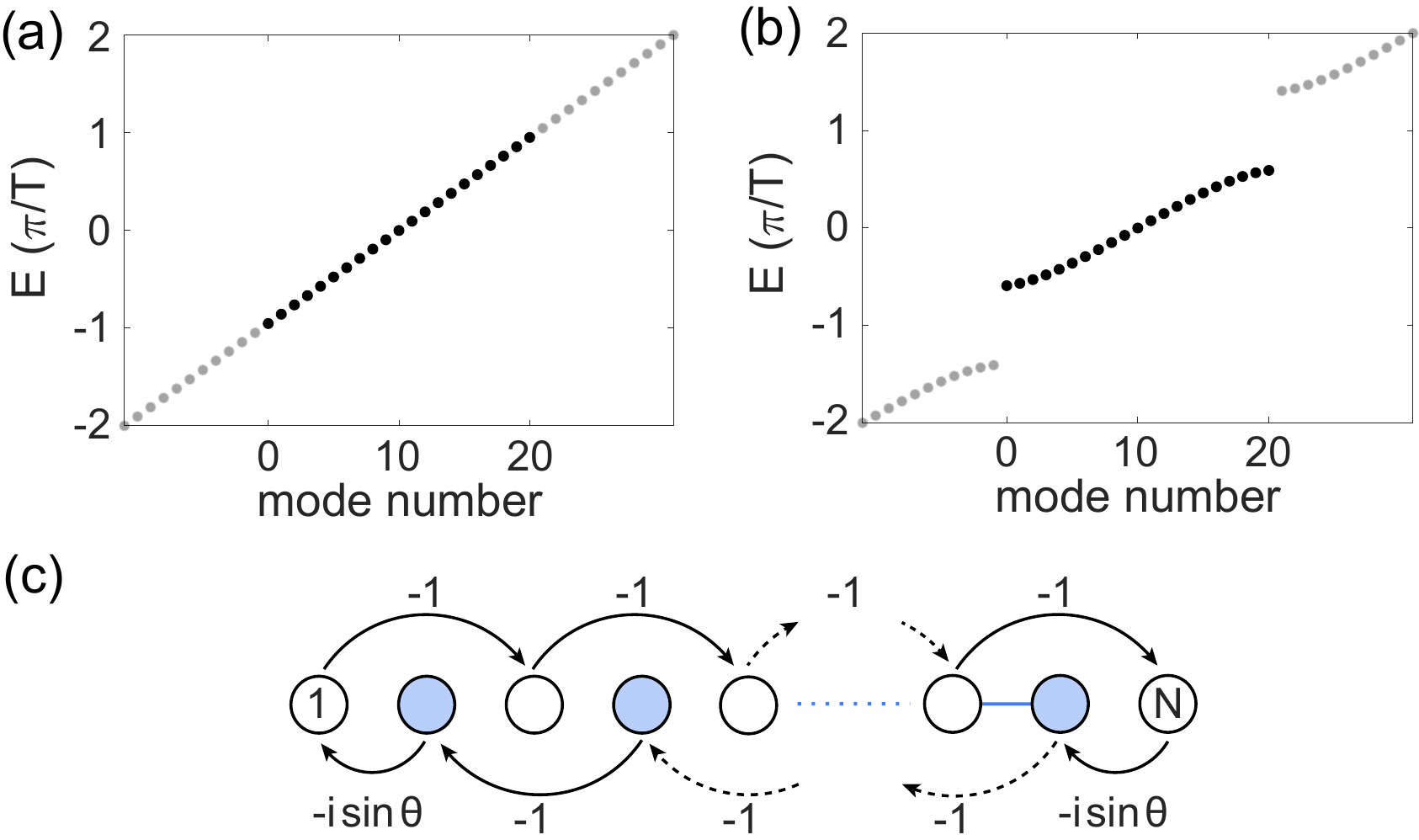}
\caption{(a,b) Same as Fig.~\ref{fig:1}(c) but with a finite lattice with $N=21$ sites and $T/\pi=1,0.6$, respectively. (c) Itinerary of a particle moving on the lattice with an odd number of sites after each Floquet period.  
}
\label{fig:2}
\end{figure}

When this system is finite with $N$ sites, the lattice momentum $k$ is no longer a good quantum number. Nevertheless, we observe an equally spaced energy ladder with $N$ levels when $\cos\theta=0$ [see Fig.~\ref{fig:2}(a)], corresponding to the linear dispersion of the Dirac cone. 

To relate this energy ladder and the WH group, we first assume an odd [$N=2m+1\,(m\in\mathbb{Z})$] number of sites (see End Matter for the case with an even $N$) and write $h_1,h_2$ in the sublattice basis, where the odd-numbered lattice sites (i.e., the A sublattice) appear first:
\begin{equation}
    h_1 = g
    \begin{pmatrix}
    0 & C \\
    C^T & 0 
    \end{pmatrix},\quad
    h_2 = g
    \begin{pmatrix}
    0 & D \\
    D^T & 0 
    \end{pmatrix},\label{eq:H1H3}
\end{equation}
Here $C$ and $D$ are $(m+1)\times m$ matrices given by
$C_{nn'} = \delta_{nn'}$ and $D_{nn'} = \delta_{n,n'+1}$. Denoting $\bm{1}_m$ as the identity matrix with $m$ rows, we find  
\be
    h_1^2 = g^2 
    \begin{pmatrix}
    \bm{1}_m &   & \\
                          & 0 & \\
                          &   & \bm{1}_m \\
    \end{pmatrix}\equiv g^2\bar{\bm{1}},\,
    h_2^2 = g^2 
    \begin{pmatrix}
    0 & \\
      & \bm{1}_{N-1} \\
    \end{pmatrix}\equiv g^2\underline{\bm{1}},\nonumber
\ee
and $h_j^3 = g^2 h_j\,(j=1,2)$, which lead to
\begin{align}
    U_1 &= \bm{1} + (\cos\theta-1)\bar{\bm{1}} - i(\sin\theta)h_1/g,\\
    U_2 &= \bm{1} + (\cos\theta-1)\underline{\bm{1}} - i(\sin\theta)h_2/g,
\end{align}
where $\theta$ again equals $gT/2$. 
Now with $\cos \theta=0$ which leads to the Dirac cone in the infinite lattice, the temporal evolution operator $U(T)$ after a Floquet period becomes 
\be 
U(T) = U_2U_1 = - iS\sin\theta  - 
\begin{pmatrix}
DC^T & \\
& D^TC
\end{pmatrix},
\ee 
where $\sin\theta$ can be either 1 or $-1$. $S$ is a sparse matrix with only two non-zero elements, i.e., the $(m+2)$th element in the first row and the $(m+1)$th element in the last row, both of which are 1. $DC^T$ is an $(m+1)\times(m+1)$ matrix, with 1's on the first lower diagonal and zeros elsewhere. Similarly, $D^TC$ is an $m\times m$ matrix, with 1's on the first upper diagonal and zeros elsewhere. 

In other words, $U(T)$ is a permutation matrix with additional phases [see the itinerary in Fig.~\ref{fig:2}(c)], which resembles Thouless pumping but without the adiabatic requirement \cite{jurgensen_quantized_2021}.
Observing that the total phase factor acquired after a round trip is $(-1)^{2m-1}(-i)^2=1$, we perform a gauge transformation on the wave function at each lattice site such that all non-zero elements of $U(T)$ become 1 (see End Matter). Furthermore, by rearranging the sites in the order of the itinerary shown in Fig.~\ref{fig:2}(c), $U(T)$ takes the form of the permutation matrix given by Eq.~(\ref{eq:Uperm}) in this ``itinerant basis.'' These transformations do not change the energy levels, which are given simply by $n\pi/NT\,(n=0,1,\ldots,2m-1)$ in our previous energy convention and $n=-m,\ldots,m$ in the first Floquet BZ shown in Fig.~\ref{fig:2}(a).

To confirm the existence of the WH group and the diagonal cyclic ladder operators $d,d^\dagger$ given by Eq.~(\ref{eq:a_altBasis}), we use the itinerant basis and note that the energy eigenstates are also the eigenstates of $U(T)$. In other words, they have the same forms as those given by Eq.~(\ref{eq:planewave}) due to the circulant form of $U(T)$. If $d$ is diagonal, then $d|\psi_n\rangle = |\psi_{n-1}\rangle$ implies that the $j$th diagonal element of $d$ is found by the ratio of the energy eigenstates $|\psi_{n-1}\rangle$ and $|\psi_{n}\rangle\,(n>0)$ at site $j$, which should not depend on the mode index $n$. We confirm this property using Eq.~(\ref{eq:planewave}), i.e.,
\be
d_{j,k} = \delta_{j,k} \left[1,\frac{\nu^{n-1}}{\nu^{n}},\frac{\nu^{2(n-1)}}{\nu^{2n}},\ldots,\frac{\nu^{(N-1)(n-1)}}{\nu^{(N-1)n}}\right]  
\ee
which is the form given by Eq.~(\ref{eq:a_altBasis}). The cyclic property of $d$ can be verified by taking the same ratios between $|\psi_{0}\rangle$ and $|\psi_{N-1}\rangle$, which gives the same expression for $d$ as above.
This approach to obtaining $d$ (and $d^\dagger$) also shows that they are gauge invariant: Adding a site-dependent phase to all the wave functions does not change these diagonal forms of the cyclic ladder operators.



In summary, we introduced cyclic ladder operators in both Hermitian and non-Hermitian systems with a finite Hilbert space, and we showed that an equally spaced energy ladder emerges when these systems have an underlying WH commutation relation, with the cyclic ladder operators and the temporal evolution operator behaving as the WH group generators. We further illustrated that such a system can be realized using a one-dimensional Floquet lattice, where the cyclic ladder operators become diagonal and the temporal evolution simplifies to a permutation matrix after a Floquet period. Our findings reveal a hidden relation between non-trivial dynamics and algebraic principles in Floquet systems, which may exist for other quantum numbers as well besides the energy levels.\\

\begin{acknowledgments}
We thank Sriram Ganeshan and Konstantinos Makris for helpful discussions. This project is supported by the National Science Foundation (NSF) under Grant No. DMR-2326698.
\end{acknowledgments}

\section{End Matter}

\subsection{WH commutation relation}

To derive Eq.~(\ref{eq:WH2}), we first prove that 
\be 
u_1U(t)a = aU(t) + \frac{1-u_N}{N}b\label{eq:WH3}
\ee
holds at any time $t$, where $u_n = e^{-in\Delta t}$. We then recover Eq.~(\ref{eq:WH2}) when $u_N=1$, i.e., at $t=2\pi/N\Delta\equiv T$. 

In our proof, we first note that
\be
u_1U(t) = e^{-i(H+\Delta)t}.
\ee
We then expand it and $U(t),u_N$ in terms of $t$, and Eq.~(\ref{eq:WH3}) becomes:
\be
\sum_n \frac{[-i(H+\Delta)t]^n}{n!}a = \sum_n \frac{a(-iHt)^n-\frac{b}{N}(-iN\Delta t)^n}{n!} + \frac{b}{N}.\label{eq:exp}
\ee
For it to hold at any time, the coefficients of any order of $t$ on the two sides must be equal to each other. For $t$-independent terms, we simply find $a=a$. For terms proportional to $t$, we require
\be 
(H+\Delta)a = aH - \Delta b,\label{eq:itr1}
\ee
which is warranted by the commutation relation between the Hamiltonian and $a$, i.e., the first relation in Eq.~(\ref{eq:cLadder2}).
For higher-order terms, we also utilize the products of $H$ and $b$:
\be
bH = 0,\quad Hb = (N-1)\Delta b.\label{eq:bH}
\ee
We rewrite the second relation as
\be
(H+\Delta) b = Nb\label{eq:Hb}
\ee
for convenience in the next step. Assuming that we have obtained
\be
(H+\Delta)^n a= aH^n - \Delta N^{n-1}b\label{eq:itrn}
\ee
which is just Eq.~(\ref{eq:itr1}) when $n=1$, we multiply it by $(H+\Delta)$ from the left and find
\be
(H+\Delta)^{n+1} a= (H+\Delta)aH^n - \Delta N^{n}b,
\label{eq:itrn+1}
\ee
where we have used Eq.~(\ref{eq:Hb}) in the second term on the right. Using Eq.~(\ref{eq:itr1}) and $bH=0$ from Eq.~(\ref{eq:bH}), the first term on the right of Eq.~(\ref{eq:itrn+1}) becomes 
\be
aH^{n+1} - \Delta bH^n = aH^{n+1}.
\ee
Therefore, we find that Eq.~(\ref{eq:itrn}) holds for $n+1$ as well. This iteration continues indefinitely, with which we conclude the proof of the expansion (\ref{eq:exp}), and in turn, the relation (\ref{eq:WH3}).

\subsection{Floquet lattices}

In the main text, we derived the analytical forms of the temporal evolution operation $U(T)$ after a Floquet period in a one-dimensional lattice with $N=2m+1\,(m\in\mathbb{Z})$ sites. Once we put $U(T)$ in the itinerant basis, the gauge transformation mentioned in the main text is given by
\be
U_p(T) = PU(T) P^{-1}
\ee
where
\be  
P = \begin{bmatrix}
\text{Diag}(1,-1,1,-1,\ldots) & \\
& r\text{Diag}(1,-1,1,-1,\ldots)
\end{bmatrix}\nonumber
\ee  
for the two sublattices, where $r=i(-1)^m \sin\theta$ and $\theta=g T/2$. $U_p(T)$ is then the permutation matrix 
\be 
U_p = 
\begin{pmatrix}
 & 1\\ 
\bm{1}_{N-1} &
\end{pmatrix}
\ee  
where $\bm{1}_{j}$ is the identity matrix with $j$ rows. 

Here we also mention that when $N$ is even (denoted by $2m$), there is still a single element of modulus 1 in each row and column of $U(T)$ before the gauge transformation [with two $-i$'s or $i$'s and $2(m-1)$ negative ones]. Therefore, the phase factor after a round trip on the lattice is $-1$ instead of 1, and we can gauge transform $U(T)$ to
\be 
U_p(T) = 
\begin{pmatrix}
& -1 \\
\bm{1}_{N-1} & 
\end{pmatrix}
\ee  
using the same form of $P$ above but with $r=i(-1)^{m+1} \sin\theta$. It can then be easily shown that the WH commutation relation (\ref{eq:WH2}) still holds in the $N$-even case, where the cyclic ladder operator is again diagonal and given by Eq.~(\ref{eq:a_altBasis}).

We also note that the determinant of $U_p(T)$ is still 1, and its eigenvalues are still equally spaced on the unit circle in the complex plane, satisfying $\lambda_n=\lambda_{N-n-1}^*$:
\be
\lambda_n = \nu^{-(n+\frac{1}{2})} \;\;(n=0,1,\ldots,N-1).
\ee
The corresponding quasi-energy ladder of $H_\text{eff}$ is then given by
\be
E_n = \frac{\pi(2n+1)}{NT},
\ee
which is again equally spaced. Because $E_0$ is no longer 0 but at $\Delta/2$, we find  
that $U(T)^N = U(NT)=-1$ instead of 1. Consequently, $2NT$ is the time needed to recover the original wave function, with $U(T)^{2N} = U(2NT)=1$.



\bibliography{references2}

\end{document}